\documentstyle[honnef, twoside]{article}

\input{psfig.sty}

\def\Journal#1#2#3#4{(#1) {#2} {\bf #3}, #4}

\def\AAp{\em Astron. Astrophys.}
\def\AApS{\em Astron. Astrophys. Suppl.}
\def\AJ{\em Astron.~J.}
\def\ApJ{\em Astrophys.~J.}

\def\ApJS{\em Astrophys.~J.~Suppl.}

\def\ARAaAp{\em Annu. Rev. Astron. Astrophys.}

\def\Nat{\em Nature\/}

\newcommand{\HII}{{\rm H\,\scriptstyle II}}
\newcommand{\cmcube}{\,{\rm cm^{-3}}}

\def\muG{\,\mu{\rm G}}
\def\radm{\,{\rm rad/m^2}}
\def\pheins{\phantom{1}}
\begin{document}

\markboth{Rainer Beck}{Radio Continuum Emission from M\,31 and M\,33}

\thispagestyle{plain}

\title{Radio Continuum Emission from M\,31 and M\,33}

\author{Rainer Beck}

\address{Max-Planck-Institut f\"ur Radioastronomie,\\
Auf dem H\"ugel 69, 53121 Bonn, Germany}

\maketitle

\abstract{The radio emission from M\,31 (like HI, CO, FIR and H$\alpha$) is
concentrated in the ``10~kpc ring'',  giving an impressive example
that cosmic rays are produced in star-forming regions.
M\,31 and M\,33 have similar strengths of the total magnetic field, but very
different field structures: The field structure in M\,31 is exceptionally
regular while that in M\,33 is rather irregular compared with other
spiral galaxies. In M\,33 the polarized intensity is highest between the
spiral arms, similar to most spiral galaxies, while in M\,31 total and
polarized emission both emerge from the ring. Star formation
in M\,31 is probably too weak to tangle the regular field.
The high regularity of the field in the M\,31 ring allows fast cosmic-ray
propagation. As a consequence, there is {\it no equipartition\/} between
the energy densities of cosmic rays and total magnetic fields.
Faraday rotation measures show that the regular field in the ring is
uni-directional, signature of the basic {\it axisymmetric} dynamo mode
with a pitch angle of only $-12\degr$.
Faraday rotation of polarized background sources shows that the
regular field and thermal gas in M\,31 extend to at least 25~kpc radius.
The regular field in M\,33 forms an open spiral, a mixture of axisymmetric
and higher modes, with the largest pitch angle ($\simeq 60\degr$) observed
in any spiral galaxy so far.
{\it Vertical filaments} in the NW and SE indicate interaction between
the thin and the thick disk of M\,31. The total emission in
the {\it central region} of M\,31 follows the spiral and radial H$\alpha$
filaments, while the polarized emission is strongest on the inner edge of
the southern spiral filament. }

\section{Introduction}

M\,31 and M\,33 impress by their enormous angular extent on sky.
Even single-dish radio telescopes with low angular resolution
reveal many details. M\,31 was the first spiral galaxy to be
detected in radio continuum 50 years ago (just a few months before I was
born) and the favourite galaxy for the Effelsberg 100-m telescope (see
Wielebinski, this volume). The ring-like structure seen in most
spectral ranges (except in the optical, unfortunately) gives M\,31 its
beauty. The radio ring of M\,31 (Fig.~1) was elected worth to be
displayed on a stamp of the German postal services in 1999 as an
example of ``fascinating phenomena in the Cosmos'' (see back cover).

\section{Global properties of M\,31 and M\,33}

\subsection{Radio continuum spectra}

\begin{table}[htb]
\caption{Radio continuum surveys of M\,31 with resolutions $\le5\arcmin$}
\begin{center}
\begin{tabular}{llccl} \noalign{\medskip}
\hline\noalign{\smallskip}
$\lambda$ &Telescope &Resolution &rms noise  &Reference \\
(cm)      &          &           &(mJy/beam) \\
\noalign{\smallskip}
\hline\noalign{\smallskip}
91.6   &Westerbork  &$53\arcsec\times 81\arcsec$ &1
    &Golla (1989)\\
73.5   &Cambridge+Effelsberg &$4\arcmin/5\arcmin$ &10 & Pooley (1969), Beck \& Gr\"ave (1982) \\
49.1   &Westerbork  &$54\arcsec\times 82\arcsec$ &0.7
    &Bystedt et al. (1984) \\
21.2   &Westerbork  &$23\arcsec\times 35\arcsec$ &0.2
    &Walterbos et al. (1985)\\
20.5   &VLA (N only)    &$5\arcsec$ &0.03 &Braun (1990) \\
20.5$^*$ &VLA+Effelsberg &$45\arcsec$ &0.05 &Beck et al. (1998) \\
11.1   &Effelsberg  &$4\farcm 8$  &2  &Berkhuijsen \&
    Wielebinski (1974) \\
11.1$^*$  &Effelsberg  &$4\farcm 4$  &1.5  &Beck et al. (1980) \\
\pheins 6.3    &Effelsberg  &$2\farcm 6$  &1.4  &Berkhuijsen et al.
    (1983) \\
\pheins 6.2$^*$  &Effelsberg  &$2\farcm 4$  &0.6 &Berkhuijsen et al.
    (in prep.) \\
\noalign{\smallskip}
\hline
\noalign{\smallskip}
\multicolumn{5}{l}{$^*$including polarization} \\
\end{tabular}
\end{center}
\end{table}

\begin{table}[htb]
\caption{Radio continuum surveys of M\,33 beyond 800~MHz}
\begin{center}
\begin{tabular}{llccl} \noalign{\medskip}
\hline\noalign{\smallskip}
$\lambda$ &Telescope &Resolution &rms noise &Reference \\
(cm)      &          &           &(mJy/beam) \\
\noalign{\smallskip}
\hline\noalign{\smallskip}
35.6   &Effelsberg  &$15\arcmin$ &$\sim$50  &Beck (1979) \\
21.1$^*$  &Effelsberg &$9\farcm 2$ &17 &Buczilowski \& Beck (1987) \\
21.1   &VLA+Westerbork  &$7\arcsec$ &0.05 &Duric et al. (1993) \\
17.4$^*$  &Effelsberg  &$7\farcm 7$ &9 &Buczilowski \& Beck (1987) \\
11.1$^*$  &Effelsberg  &$4\farcm 4$ &3 &Buczilowski \& Beck (1987) \\
11.1$^*$  &Effelsberg  &$4\farcm 4$ &0.6 &Beck (unpubl.) \\
\pheins 6.3$^*$   &Effelsberg  &$2\farcm 4$  &2  &Buczilowski \& Beck
    (1987) \\
\pheins 6.2$^*$   &Effelsberg  &$2\farcm 4$  &0.6 &Niklas \& Beck
    (unpubl.) \\
\pheins 6.2    &VLA+Westerbork &$7\arcsec$  &0.05 &Duric et al. (1993) \\
\pheins 2.8$^*$  &Effelsberg  &$1\farcm 2$  &1.8 &Buczilowski \& Beck
    (1987) \\
\noalign{\smallskip}
\hline
\noalign{\smallskip}
\multicolumn{5}{l}{$^*$including polarization} \\
\end{tabular}
\end{center}
\end{table}

Tables~1 and 2 give the radio continuum surveys of M\,31 and M\,33 with
sufficiently small telescope beams to resolve the spiral arms.
The flux densities (integrated to 16~kpc and 12~kpc radius,
respectively) give an average spectral index in M\,31 and M\,33 of
$0.76 \pm 0.04$ and $0.75 \pm 0.10$, respectively.
The spectrum of M\,33 flattens below $\simeq800$~MHz due to thermal
absorption (Israel et al., 1992).
A nonthermal (synchrotron) spectral index of
$\alpha_{\rm nt}$ ($\alpha_{\rm nt} \simeq -1.0$ for M\,31 and
$\alpha_{\rm nt} \simeq -0.9$ for M\,33) was derived from spectral index
maps in regions with low star-forming activity where the thermal
contribution is negligible. Assuming that $\alpha_{\rm nt}$ is constant
across the galaxy, the nonthermal and thermal components were separated
(Berkhuijsen et al., in prep). The average thermal fractions at $\lambda6$~cm
are $\simeq 35\%$ and $\simeq 30\%$ in M\,31 and M\,33, respectively.
The large value in M\,31 is surprising in view of its relatively low
star-formation rate.

The radial distributions of the total radio continuum emission from
the diffuse disk of M\,33 (Fig.~6) and beyond the ``ring'' of M\,31 (Fig.~1)
can be fitted well by exponentials (Table~3).

\subsection{Magnetic field strengths}

The average and maximum nonthermal and polarized intensities
at $\lambda6$~cm and at $3\arcmin$
resolution were used to compute the average and maximum equipartition
strengths of the total field and its regular component, assuming a ratio
of cosmic-ray protons to electrons of 100 and pathlengths through the
emitting regions of 3~kpc and 2~kpc (taking into account the different
inclinations). Equipartition is probably valid on average over the ring
of M\,31 but invalid within the ring (Sect.~3.2), so the maximum field
strengths are lower limits.
The results are summarized in Table~3. The two galaxies are
similar in most of their radio continuum properties.
The main differences between M\,31 and M\,33 are the morphology (the ``ring''
in M\,31, a diffuse disk in M\,33) and the degree of regularity of the
magnetic field which is much higher in M\,31.
Corrected for thermal contribution, the nonthermal degree
of polarization at $\lambda6$~cm at $3\arcmin$ resolution becomes 33\%
in M\,31, only 10\% in M\,33. No other spiral galaxy reveals a degree of
polarization as high as M\,31, even when observed with higher spatial
resolution.

\begin{table}[htb]
\caption{Average radio continuum properties}
\begin{center}
\begin{tabular}{lr@{$\pm$}lr@{$\pm$}l}
\noalign{\medskip}
\hline\noalign{\smallskip}
    &\multicolumn{2}{c}{M\,31} &\multicolumn{2}{c}{M\,33} \\
\noalign{\smallskip}
\hline\noalign{\smallskip}
Spectral index of the integrated emission &0.76 &0.04 &0.75 &0.10 \\
Nonthermal spectral index                 &1.0 &0.1   &0.9 &0.1 \\
Thermal fraction at $\lambda$6~cm  &\multicolumn{2}{c}{$\simeq\,
    35\%$} &\multicolumn{2}{c}{$\simeq\, 30\%$} \\
Average total field strength  &6 & 2 $\mu$G  &6 & 2 $\mu$G \\
Maximum total field strength  &$\ge\, 8$ & 2 $\mu$G &9 & 2 $\mu$G \\
Average regular field strength &4 &1 $\mu$G &2 &1 $\mu$G \\
Maximum regular field strength &$\ge\, 6$ &1 $\mu$G &4 &1 $\mu$G \\
Scale length of the disk at $\lambda$6~cm &3.3 &0.1 kpc$^*$ &2.3
    &0.1 kpc \\
\noalign{\smallskip}
\hline
\noalign{\smallskip}
\multicolumn{4}{l}{$^*$beyond 10 kpc radius} \\
\end{tabular}
\end{center}
\end{table}

\section{The 10~kpc ``ring'' of M\,31}

\subsection{Magnetic field structure}

Radio continuum emission tells us about the distribution of
magnetic fields and cosmic rays. Figure~1 shows the total and
polarized intensities of M\,31 at $\lambda6$~cm. The ``ring'' at about
10~kpc radius is well defined in both components. In total emission the
similarity to the ring of CO emission (Nieten et al., this volume)
and dust emission (Haas, this volume) is
striking and gives further evidence that magnetic fields are anchored
in gas clouds (Berkhuijsen et al., 1993).

The general coincidence of total and polarized emission regions in the ring
is unusual. In all other spiral galaxies observed so far, polarized intensity
is strongest between the spiral arms as traced by total emission, but
weak in the arms due to field tangling by star-formation activity
(Beck et al., 1996). Star formation activity is probably too weak to
tangle M\,31's regular field significantly.

\begin{figure}[htb]
\centerline{\psfig{figure=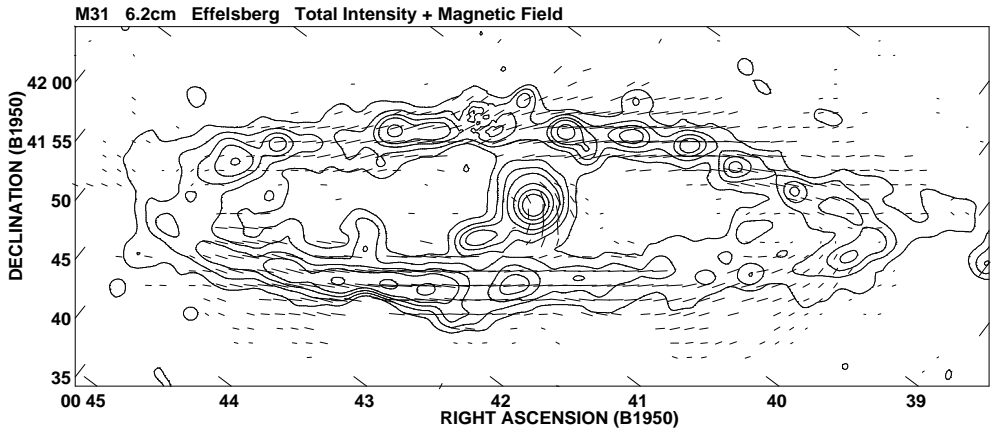,width=13truecm,angle=270,%
       bbllx=133pt,bblly=44pt,bburx=441pt,bbury=749pt}}
\caption{Total intensity of M\,31 at $\lambda6.2$~cm, observed with the
Effelsberg 100-m telescope, smoothed to $3\arcmin$ beamsize. The
lengths of the vectors are proportional to the polarized intensities, their
orientations have been corrected for Faraday rotation by using the
$\lambda11.1$~cm Effelsberg data at $5\arcmin$ resolution (Berkhuijsen
et al., in prep.)}
\end{figure}

The variations of total nonthermal intensity (corrected for thermal
emission) and polarized intensity with azimuthal angle (Fig.~2) reveal
peaks near the minor axis and minima near the major axis, while the
unpolarized nonthermal intensity shows almost no azimuthal variation.
In polarized intensity, the peaks are roughly symmetric and the minima
are deep, as expected from polarized emission from a strongly inclined
toroidal field. Polarized synchrotron emission traces the
component of the regular field {\it perpendicular to the line of sight}
which is largest near the minor axis and smallest near the major axis
of the galaxy's projected disk. Total synchrotron emission also depends on
irregular fields which lead to unpolarized emission filling the minima
of polarized emission near the major axis.

Polarized intensity varies as:
$$I_{\rm p} \propto N_{\rm CRE} \, B_{\rm reg}^{1-\alpha_{\rm nt}} \,
\sin^{1-\alpha_{\rm nt}}\phi $$
where $N_{\rm CRE}$ is the number density of cosmic-ray electrons in the
relevant energy range, $\alpha_{\rm nt}$ is the nonthermal spectral
index and $\phi$ is the viewing angle between the field and the line of
sight. In case of equipartition between cosmic rays and magnetic fields,
$N_{\rm CRE} \propto B_{\rm tot}^2$.

For an axisymmetric spiral field with a pitch angle $p$:
$$ \cos\phi =  \cos(\Theta - p) \, \sin i $$
where $\Theta$ is the azimuthal angle in the plane of the galaxy
and $i$ is the galaxy's inclination. For
M\,31, $\alpha_{\rm nt} \simeq -1$ (see above), $p \simeq -12\degr$
(Fletcher et al., this volume) and $i \simeq 78\degr$.

The variation of $\ln I_{\rm p}$ with $\ln | \sin\phi |$ was computed
between 8 and 12~kpc radius\footnote{based on the ``old'' distance of
690~kpc, for better comparison with previous results}.
In three quadrants the slopes are around 2, a flatter slope
of 1.3 was found only in the NW quadrant. The slope
of $\simeq 2$ means that the variation of $I_{\rm p}$ can be described by the
variation in $\phi$ due to geometry while $N_{\rm CRE}$ and $B_{\rm reg}$
are roughly constant along the ring.
Locally in the ring, the nonthermal degree of polarization
increases to $\simeq 50\%$.
{\it The magnetic field of M\,31 is exceptionally regular.}

\smallskip

\begin{figure}[htb]
\hbox to \hsize{
\psfig{figure=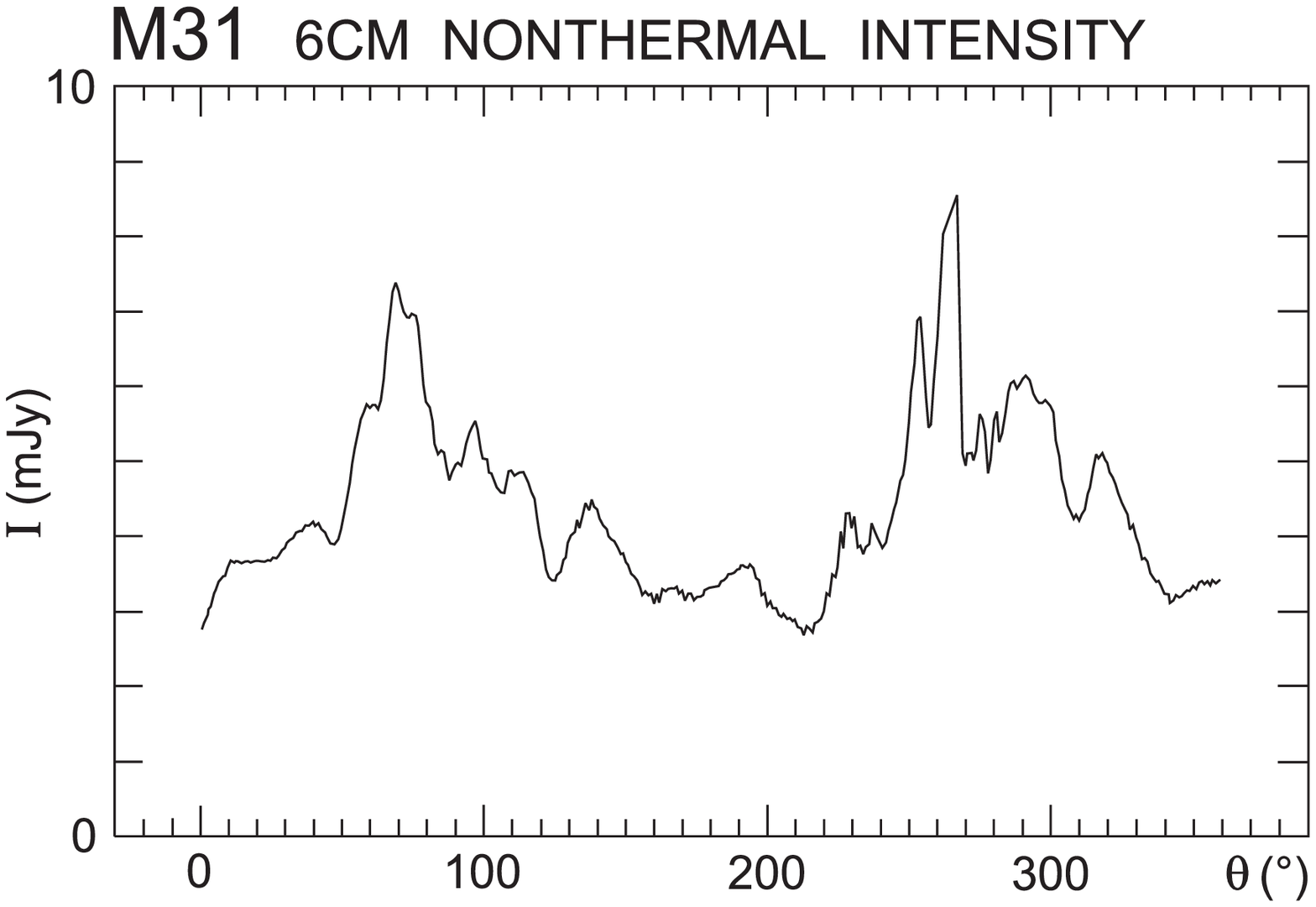,width=7truecm,%
       bbllx=169pt,bblly=132pt,bburx=667pt,bbury=473pt}
\hfill
\psfig{figure=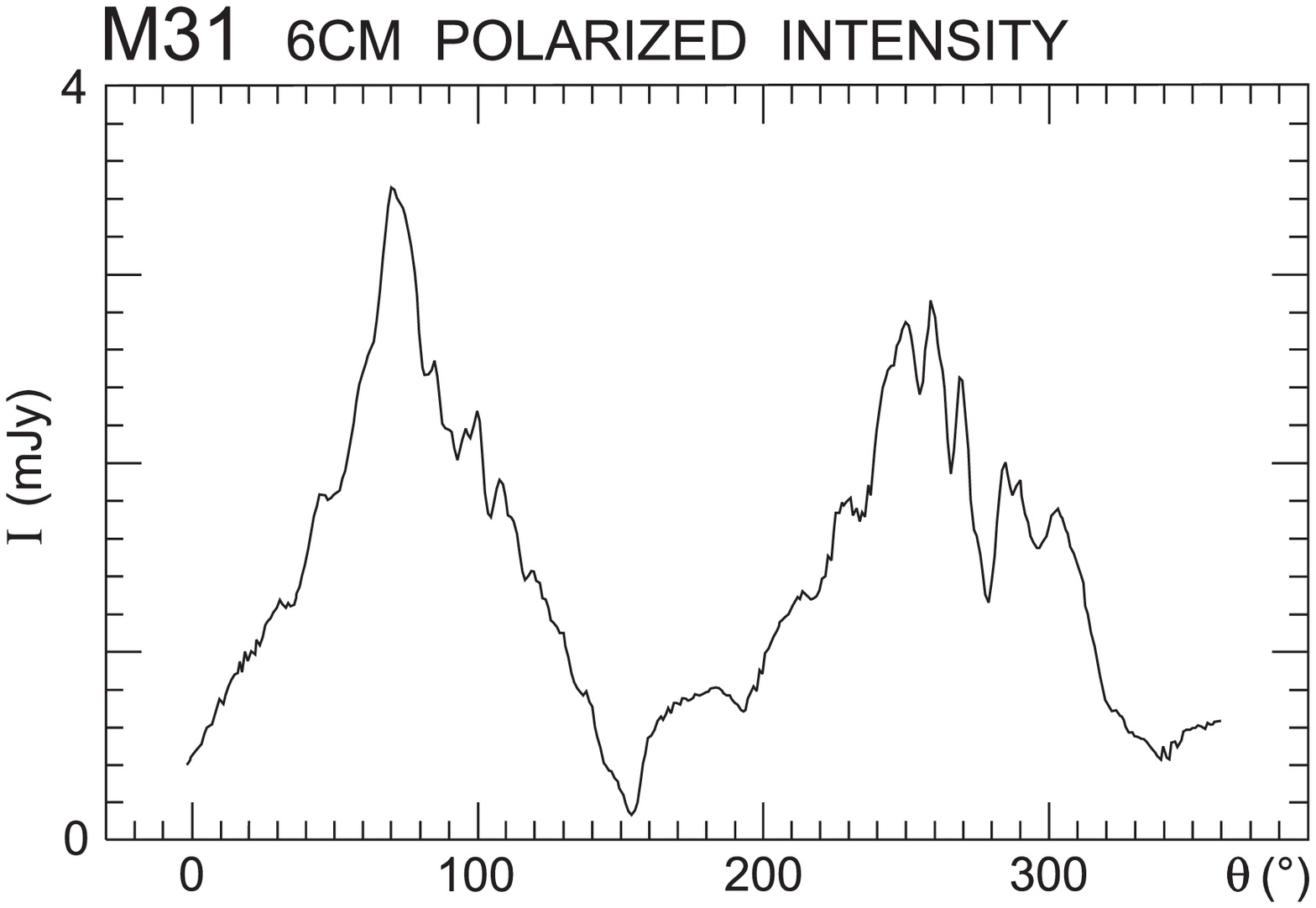,width=7truecm,%
     bbllx=175pt,bblly=130pt,bburx=669pt,bbury=470pt}}
\caption{Variation of nonthermal intensity (a) and polarized intensity
(b) at $\lambda6.2$~cm ($3\arcmin$ resolution) between 8 and 12~kpc
radius with azimuthal angle, counted counter-clockwise in the plane of M\,31,
starting from the northeastern major axis}
\end{figure}

\subsection{Cosmic rays}

The total nonthermal continuum intensity (the signature of $B_{\rm tot}^2$)
at $\lambda6$~cm shows a much weaker variation with viewing angle $\phi$
than the polarized intensity (slopes of 0.3 -- 0.5 compared with
$\simeq 2$).
In case of equipartition between cosmic rays and magnetic fields,
$I_{\rm p}$ also depends on $B_{\rm tot}^2$ (see above) which should
flatten the variation of $I_{\rm p}$ with viewing angle, in contrast
to observations. Cosmic-ray energy density is {\it not in equipartition
with the total magnetic field}, but almost constant along the ring.
Urbanik et al. (1994) and Hoernes et al. (1998a) came to a similar conclusion,
based on the comparison between the radio and FIR intensities.
The regular field dominates in the ring. Cosmic rays can propagate along
the (almost) toroidal field and fill the torus smoothly, without being
scattered by field irregularities. This allows diffusion speeds
{\it larger than the Alfv\'en speed}.

As the regular fields extend to at least 25~kpc radius (Han et al.,
1998), the concentration of the radio continuum emission to the ring
is a result of the cosmic-ray distribution. Star formation in M\,31 is
mainly occuring in the ring, and the limitation of cosmic rays to
the same region is an impressive confirmation that these
are accelerated in Pop~I objects, e.g. shock fronts of supernova remnants
(Duric, this volume).
The energy density of cosmic rays drops outside of the ring (ie.
at smaller and larger radii), while the field strength is almost radially
constant (Han et al., 1998), so that the energy density of the field is
larger than that of the cosmic rays outside of the ring: energy
equipartition is also invalid outside the ring.

\subsection{Rotation measures}

Faraday rotation measures ($RMs$) are proportional to the component of
the regular field {\it along} the line of sight and to the density of
thermal electrons ($n_{\rm e}$). In case of a toroidal regular field,
$| RM |$ is maximum on the major axis and zero on the minor axis; $RM$
varies sinusoidally with azimuthal angle. If the regular field is of
axisymmetric spiral type (ASS), the variation of $RM$ with azimuthal
angle is phase-shifted by an angle equal to the pitch angle of the
magnetic spiral.

Figure~3 shows the $RMs$ in the M\,31 ring,
derived from the Effelsberg data at $\lambda6$~cm and
$\lambda11$~cm at $5\arcmin$ resolution. The average $RM$ of about
$-90\radm$ is due to the foreground medium in the Galaxy. The azimuthal
variation between 8 and 12~kpc radius can be fitted by a sine wave,
the signature of an axisymmetric (ASS) field as expected from dynamo
models (Shukurov, this volume). A detailed analysis of all available
polarization data including, those at $\lambda20$~cm, is given by
Fletcher et al. (this volume). The large-scale pattern in the $RM$ map
is the proof that the field in M\,31 is not only regular, but also {\it coherent}
as it preserves its direction all over the galaxy. The radial component of
the field points {\it towards the center of M\,31} everywhere.

\begin{figure}[htb]
\centerline{\psfig{figure=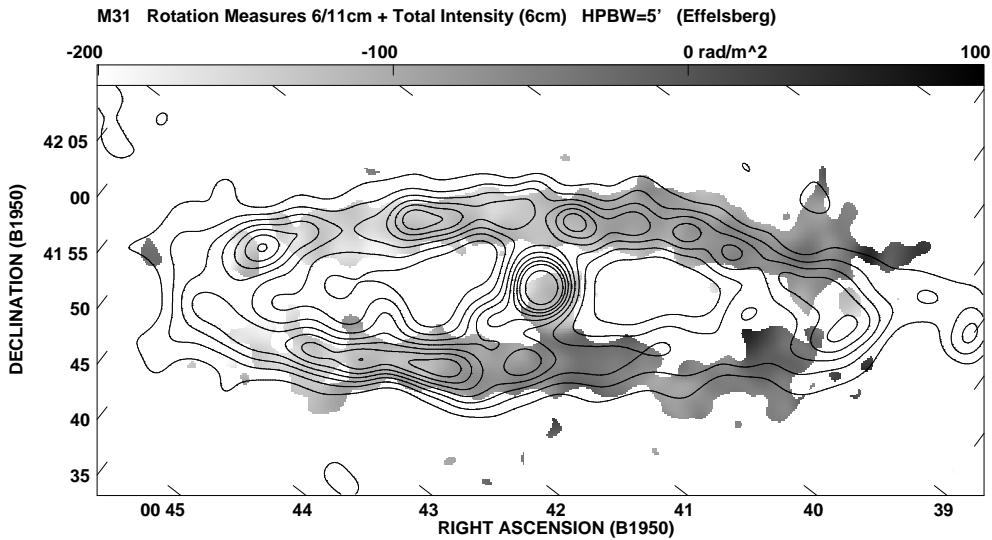,width=13truecm,angle=270,%
       bbllx=89pt,bblly=39pt,bburx=473pt,bbury=750pt}}
\caption{Faraday rotation measures between $\lambda6.2$~cm and
$\lambda11.1$~cm at $5\arcmin$ resolution, computed in regions where
the polarized intensities at both wavelengths are larger than 5$\times$
the rms noise (Berkhuijsen et al., in prep.)}
\end{figure}

The $RM$ distribution in Fig.~3 is much smoother than that of the
thermal emission at the same resolution. While the latter is
sensitive to $n_{\rm e}^2$ and thus mainly traces peaks in thermal
density ($\HII$ regions), $RM$ traces the diffuse, extended component
of $n_{\rm e}$. The $RM$ amplitude is $77\radm$. With a regular
field strength of $4\muG$, the average electron density $n_{\rm e}$
is  $\le 0.015\cmcube$ over a pathlength of $\ge 3$~kpc.
The true extent of the diffuse thermal gas in M\,31 in unknown (Sect.~4).

\subsection{Fine structure of the field}

M\,31 was also observed at $\lambda20$~cm with the VLA with $45\arcsec$
(150~pc) resolution (Beck et al., 1998). The total continuum emission
shows remarkable asymmetries: The ring is thin (full
half-power width of 300--400~pc) in the
northeastern (Fig.~4a) and southwestern quadrants, but thick (width of
700--1000~pc) in the northwestern and
southeastern (Fig.~4b) quadrants. The latter quadrant shows several {\it
filaments} perpendicular to the ring.

\begin{figure}[htb]
\hbox to \hsize{
\psfig{figure=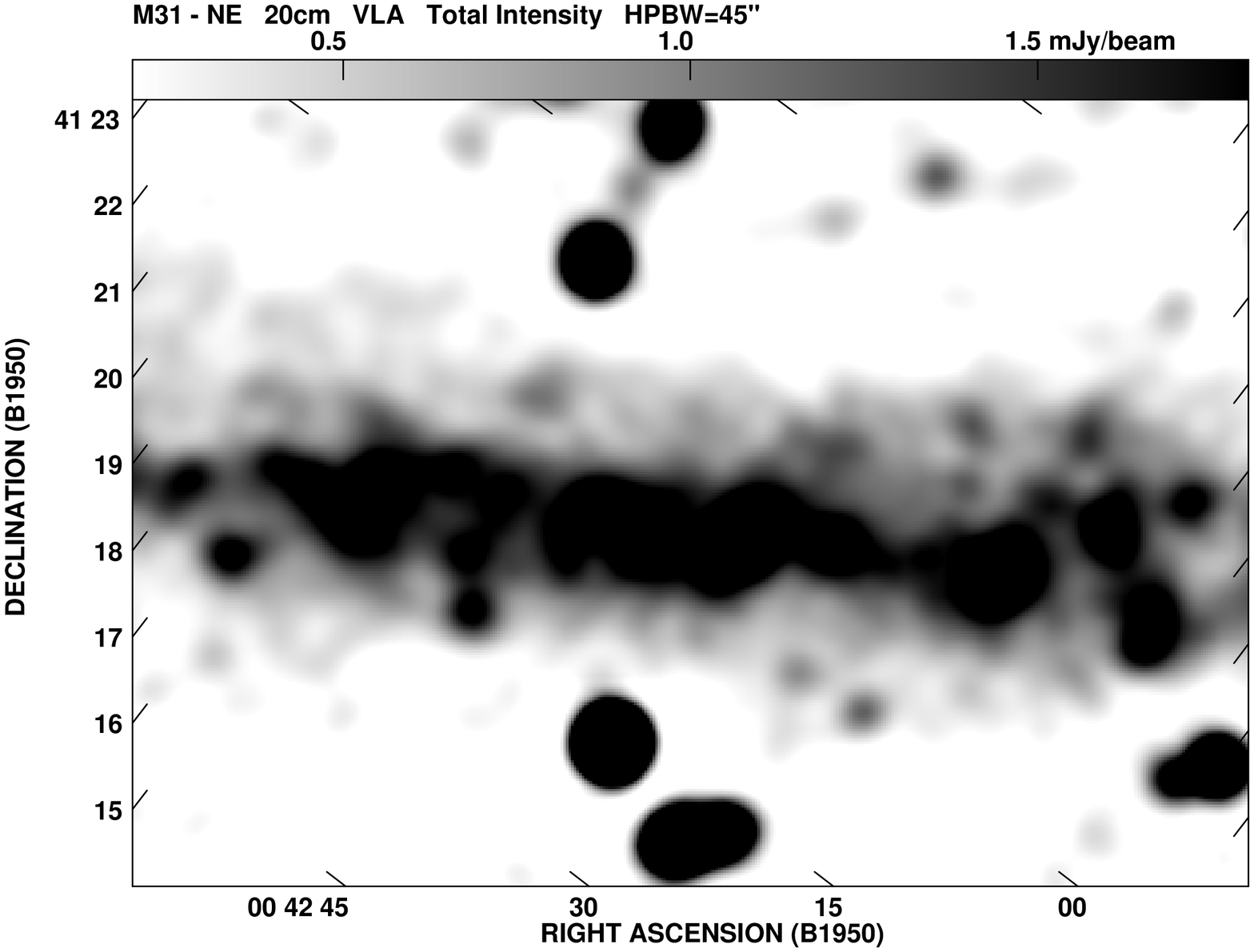,width=7.2truecm,angle=270,%
       bbllx=59pt,bblly=82pt,bburx=530pt,bbury=704pt}
\hfill
\psfig{figure=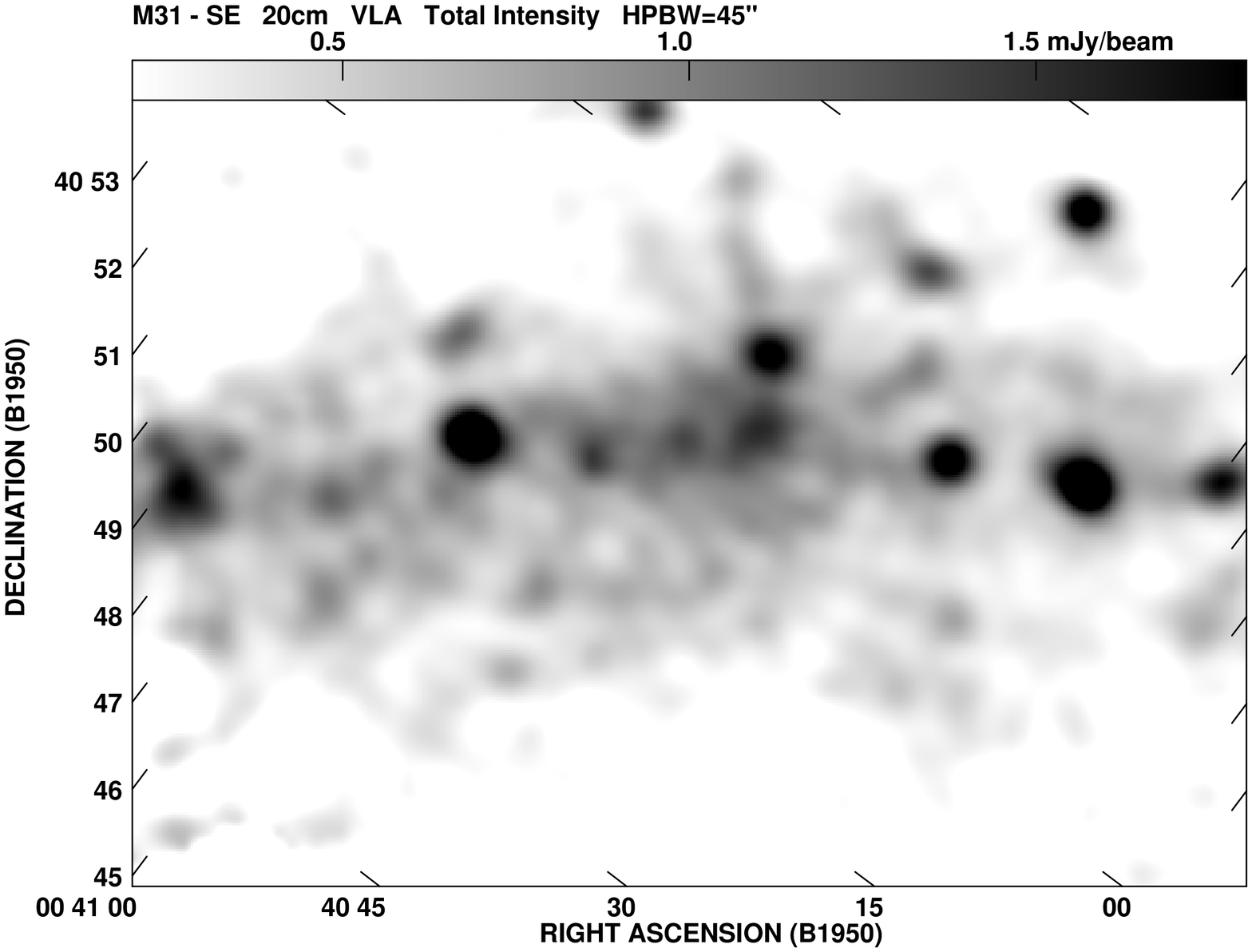,width=7.2truecm,angle=270,%
       bbllx=59pt,bblly=82pt,bburx=530pt,bbury=704pt}}
\caption{Total intensity at $\lambda20.5$~cm, observed with the VLA
with $45\arcsec$ in the eastern ring of M\,31 (Beck et al., 1998)}
\end{figure}

The highest resolution ($15\arcsec$ or 50~pc) so far was obtained with the
VLA at $\lambda6$~cm in a portion of the southwestern ring (Hoernes, 1997).
The polarization vectors show systematic fluctuations around the mean
orientation, according to Beck et al. (1989) possibly signatures of
{\it Parker instabilities}. The fine structure of the irregular field
is partly resolved. The largest scale of the turbulent field is
about 50~pc, in agreement with results from the Galaxy (e.g. Rand \&
Kulkarni, 1989).

\section{Radio halo around M\,31 ?}

Early radio observations of M\,31 indicated a radio halo around M\,31,
but with higher resolution this feature was resolved into point-like
sources and a Galactic spur emerging from the plane of the Galaxy
(Gr\"ave et al., 1981). This spur is obvious also on the deep
$\lambda20$~cm polarization map obtained with the Effelsberg telescope
(Beck et al., 1998). {\it No excess radio continuum emission is
detected around M\,31.}

On the other hand, the filaments visible in Fig.~4b indicate that some
cosmic rays and magnetic fields may be leaving the ring. Han et al.
(1998) found that the $RMs$ of polarized background sources located at
radii up to 25~kpc radius from the center of M\,31 are similar to the
$RMs$ of the emission from the ring (Fig.~3) at the
same azimuthal angle. It seems that the regular field {\it and\/} the
thermal gas extend much beyond the ring and may form a
magneto-ionic thick disk or halo. A larger sample
of $RMs$ of polarized background sources is needed to confirm
this result.

\section{The central region of M\,31}

The proximity of M\,31 enables to study its central region with high
linear resolution. Within $5\arcmin$ from the nucleus (about 1~kpc in
the plane of M\,31) deep observations in the H$\alpha$ line (Ciardullo et
al., 1988) revealed a complex structure of narrow filaments on the
northwestern side, a small bar-like structure and a spiral arm on the
southeastern side. Ciardullo et al. found evidence for an outflow of
gas from the inner 200~pc.
Radio continuum observations at $\lambda$49~cm and $\lambda$21~cm
with the Westerbork telescope showed extended emission from the
central spiral and the filaments (Walterbos \& Gr\"ave, 1985). More
recently, the central $20\arcmin$ was observed with the VLA at
$\lambda$6~cm and $\lambda$20~cm (Hoernes, 1997) with angular resolutions
of $13\arcsec$
and $22\arcsec$, respectively, corresponding to 45~pc and 150~pc in the
plane of M\,31. The distribution of the total emission at $\lambda$6~cm
is very similar to that of the H$\alpha$ emission
(Devereux et al., 1994), with enhanced emission from the spiral
arm in the SE and the brightest filaments in the NW. These results
indicate field compression by large-scale shocks, either by a wind or
by density waves.

The existence of synchrotron-emitting cosmic-ray electrons in the
central region is puzzling as star-forming activity is weak there.
Hoernes (1997) derived the distribution of the total spectral index
$\alpha$ ($S \propto \nu^{\alpha}$) between 20~cm and 6~cm with a
resolution of $22\arcsec$; it has a filamentary and patchy
appearance. At the very centre the spectrum is {\it flat} with $\alpha
\simeq -0.2$ and it steepens outwards. Hoernes (1997) showed that this
variation cannot be explained by thermal emission. Hence, the nonthermal
spectral index $\alpha_{\rm nt}$ at the center must be close to $-0.2$;
it slowly decreases along the southern arm and the filaments, but
perpendicular to these features it decreases much faster, reaching values
$\la -1.0$ at 1~kpc radius. This behaviour is similar to that observed
in the central regions of the Milky Way (Pohl et al., 1992)
and M81 (Reuter \& Lesch, 1996) and suggests the existence of a black
hole associated with a mono-energetic source of relativistic electrons
in the nucleus (Hoernes et al., 1998b).

Polarized emission at $\lambda$6~cm (Fig.~5) is weak in the NW region
of the central filaments which is probably due to strong Faraday
depolarization, but is concentrated on and around the southern
spiral arm with the highest degree of polarization along the {\it
inside\/} of the arm. This result is in favour of a density wave moving
faster than the arm and compressing the gas and field at the inner edge.

\begin{figure}[htb]
\psfig{figure=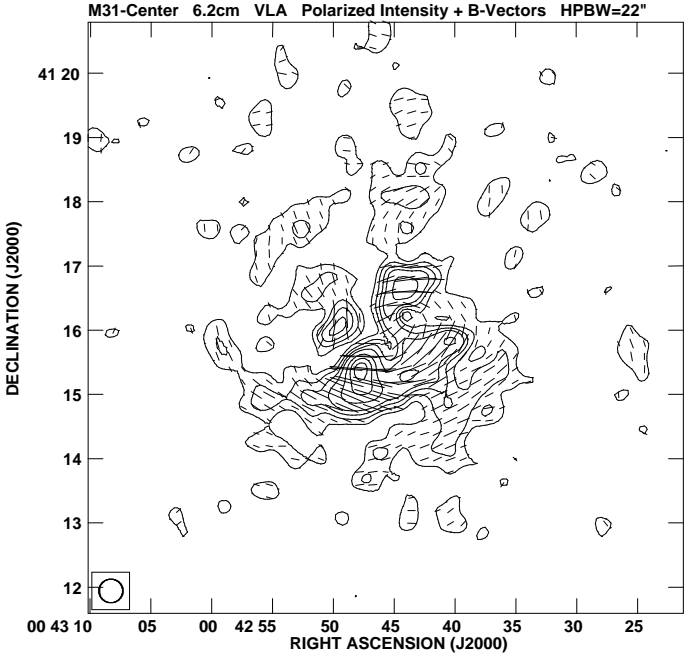,width=9truecm,%
       bbllx=39pt,bblly=160pt,bburx=567pt,bbury=670pt}
\vspace{-3truecm}
\parbox[t]{9.5cm}
\hfill
\parbox[t]{6cm}{
\caption{Polarized intensity from the central region of M\,31, observed
with the VLA and smoothed to $22\arcsec$ beamsize. The vectors lengths
are proportional to the polarized intensities, their orientations
have not been corrected for Faraday rotation (Hoernes, 1997)} }
\end{figure}

\section{The magnetic field in M\,33}

The total emission of M\,33 at $\lambda$6~cm (Fig.~6) is diffuse.
The spiral arms are hardly visible. Polarized emission
is also diffuse, but has a bright maximum {\it between\/} the two
northern spiral arms, with a maximum fractional polarization of 30\%.
M\,33 differs from M\,31 in this respect, but is in line with the regular
fields observed in the interarm regions of many other spiral galaxies
(Beck et al., 1996).

\begin{figure}[htb]
\psfig{figure=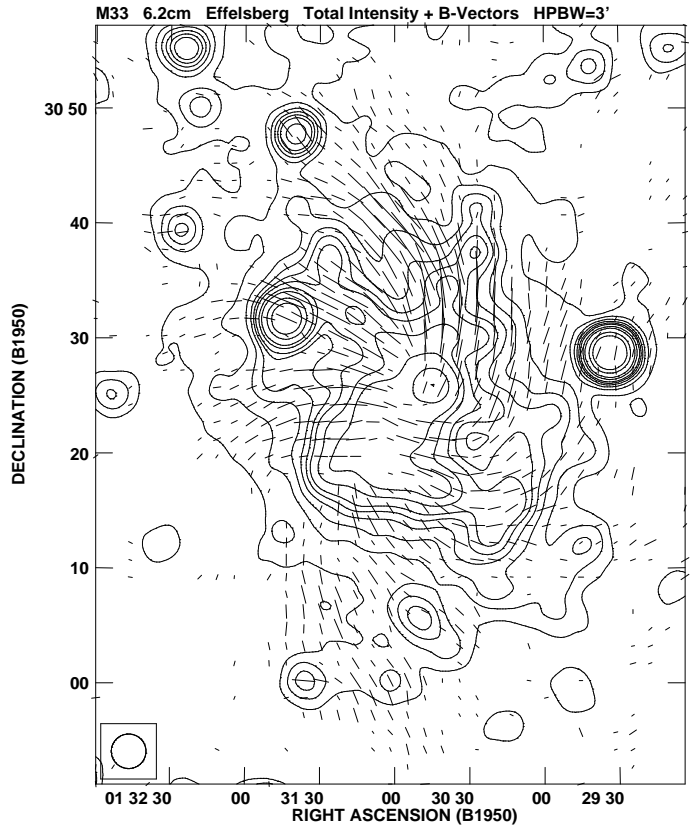,width=9truecm,%
       bbllx=46pt,bblly=108pt,bburx=562pt,bbury=735pt}
\vspace{-2.5truecm}
\parbox[t]{9.5cm}
\hfill
\parbox[t]{6cm}{
\caption{Total intensity of M\,33 at $\lambda6.2$~cm, observed with the
Effelsberg telescope, smoothed to $3\arcmin$ beamsize. The
lengths of the vectors are proportional to the polarized intensities, their
orientations have not been corrected for Faraday rotation (Niklas \&
Beck, unpubl.)} }
\end{figure}

In spite of the diffuse polarized emission, the vectors in Fig.~6
form a regular spiral pattern. The average pitch angle is
about $60\degr$, much larger than in M\,31 and the largest pitch angle
observed in any spiral galaxy so far. Faraday rotation in M\,33 is much
smaller than in M\,31 and does not reveal an obvious systematic pattern.
Detailed analysis shows that the field can be described by a mixture of
dynamo modes (Fletcher et al., this volume). The claim of a
possibly dominating biymmetric field by Buczilowski \& Beck (1991)
cannot be confirmed with the new data.

\section{Outlook}

M\,31 is an exceptional galaxy with respect to magnetic fields and
cosmic rays. The field in M\,31 is very regular
and extends far beyond the radio-emitting
regions. Surprisingly, the origin of the ``ring'' is still unexplained.

Surface brightness in M\,31 is low so that polarization
observations with higher resolution will be possible only with
next-generation radio telescopes like the {\it Square Kilometer Array}.
This will also allow to observe more $RMs$
from background sources and the detection of pulsars in M\,31.

\section*{References}\noindent

\references

Beck R. (1979) Ph.D. Thesis, University of Bonn.

Beck R. \Journal{1982}{\AAp}{106}{121}.

Beck R., Gr\"ave R. \Journal{1982}{\AAp}{105}{192}.

Beck R., Berkhuijsen E.M., Wielebinski, R.
\Journal{1980}{\Nat}{283}{272}.

Beck R., Loiseau N., Hummel E., Berkhuijsen E.M., Gr\"ave R.,
Wielebinski, R. \Journal{1989}{\AAp}{222}{58}.

Beck R., Brandenburg A., Moss D., Shukurov A., Sokoloff D.
\Journal{1996}{\ARAaAp}{34}{155}.

Beck R., Berkhuijsen E.M., Hoernes P. \Journal{1998}{\AApS}{129}{329}.

Berkhuijsen E.M., Wielebinski R. \Journal{1974}{\AAp}{34}{173}.

Berkhuijsen E.M., Wielebinski R., Beck R.
\Journal{1983}{\AAp}{117}{141}.

Berkhuijsen E.M., Bajaja E., Beck R. \Journal{1993}{\AAp}{279}{359}.

Braun R. \Journal{1990}{\ApJS}{72}{761}.

Buczilowski U.R., Beck R. \Journal{1987}{\AApS}{68}{171}.

Buczilowski U.R., Beck R. \Journal{1991}{\AAp}{241}{47}.

Bystedt J.E.V., Brinks E., de Bruyn A.G. et al.
\Journal{1984}{\AApS}{56}{245}.

Devereux N.A., Price R., Wells L.A., Duric N.
\Journal{1994}{\AJ}{108}{1664}.

Duric N., Viallefond F., Goss W.M., van der Hulst J.M.
\Journal{1993}{\AApS}{99}{217}.

Golla G. (1989) Diploma Thesis, University of Bonn.

Gr\"ave R., Emerson D.T., Wielebinski R. \Journal{1981}{\AAp}{98}{260}.

Han J.L., Beck R., Berkhuijsen E.M. \Journal{1998}{\AAp}{335}{1117}.

Hoernes P. (1997) Ph.D. Thesis, University of Bonn.

Hoernes P., Berkhuijsen E.M., Xu C. \Journal{1998a}{\AAp}{334}{57}.

Hoernes P., Beck R., Berkhuijsen E.M. (1998b) in {\em The Central
Regions of the Galaxy and Galaxies}, ed. Y. Sofue, Kluwer, Dordrecht,
p.~351.

Israel F.P., Mahoney M.J., Howarth N. \Journal{1992}{\AAp}{261}{47}.

Pohl M., Reich W., Schlickeiser R. \Journal{1992}{\AAp}{262}{448}.

Pooley, G.G. \Journal{1969}{MNRAS}{144}{101}.

Rand R.J., Kulkarni S.R. \Journal{1989}{\ApJ}{343}{760}.

Reuter H.-P., Lesch H. \Journal{1996}{\AAp}{310}{L5}.

Urbanik M., Otmianowska-Mazur K., Beck R.
\Journal{1994}{\AAp}{287}{410}.

Walterbos R.A.M., Gr\"ave R. \Journal{1985}{\AAp}{150}{L1}.

Walterbos R.A.M., Brinks E., Shane W.W. \Journal{1985}{\AApS}{61}{451}.

\end{document}